\documentclass{article}

\usepackage{arxiv}

\usepackage[utf8]{inputenc} 
\usepackage[T1]{fontenc}    
\usepackage{hyperref}       
\usepackage{url}            
\usepackage{booktabs}       
\usepackage{amsfonts}       
\usepackage{nicefrac}       
\usepackage{microtype}      
\usepackage{lipsum}		
\usepackage{amsmath} 
\usepackage{graphicx}
\usepackage{natbib}
\usepackage{doi}
\usepackage{amssymb}
\usepackage{amsmath,amssymb,amsfonts}
\usepackage{algorithmic}
\usepackage{graphicx}
\usepackage{textcomp}

\usepackage{multirow}
\usepackage{booktabs}
\usepackage{siunitx}
\title{Physics-driven Sonification for Improving Multisensory Needle Guidance in Percutaneous Epicardial Access}

\author{ \href{https://orcid.org/0009-0003-7445-2165}{\includegraphics[scale=0.06]{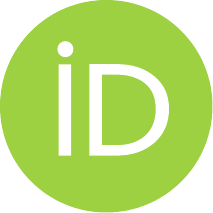}\hspace{1mm}Veronica Ruozzi\thanks{These authors contributed equally to this work.}} \\
	Department of Electronics Information and Bioengineering\\
	Politecnico di Milano\\
	Milan, Italy\\
	\And
	\href{https://orcid.org/0009-0003-6905-2001}{\includegraphics[scale=0.06]{orcid.pdf}\hspace{1mm}Sasan Matinfar\thanks{These authors contributed equally to this work.}} \\
	Chair of Computer Aided Medical Procedures\\
	Technische Universität München\\
	München, Germany
    \And
	Pasquale Vergara\\
	Department of Advanced Biomedical Sciences\\
	Federico II University of Naples\\
	Naples, Italy
    \And
	Alessandro Albanesi\\
	Department of Electronics Information and Bioengineering\\
	Politecnico di Milano\\
	Milan, Italy\\
    \And
	Serena Dell'Aversana\\
	Department of Radiology\\
	Ospedale S. Maria Delle Grazie - ASL Napoli 2 Nord\\
	Pozzuoli, Italy\\
    \And
	Stefano Carugo\\
	Department of Clinical Sciences and Community Health\\
	University of Milan\\
	Milan, Italy\\
    \And
	Gianluigi Buccoliero\\
	ASST Bergamo Ovest\\
	Bergamo, Italy\\
    \And
    Nassir Navab\\
	Chair of Computer Aided Medical Procedures\\
	Technische Universität München\\
	München, Germany
    \And
	Alberto Redaelli\\
	Department of Electronics Information and Bioengineering\\
	Politecnico di Milano\\
	Milan, Italy\\
    \And
	Emiliano Votta\\
	Department of Electronics Information and Bioengineering\\
	Politecnico di Milano\\
	Milan, Italy\\
}

\renewcommand{\undertitle}{A preprint--This work has been submitted to the IEEE for possible publication. Copyright may be transferred without notice, after which this version may no longer be accessible.}
\renewcommand{\shorttitle}{Preprint}

\hypersetup{
pdftitle={A template for the arxiv style},
pdfsubject={q-bio.NC, q-bio.QM},
pdfauthor={Ruozzi V., Matinfar S.},
pdfkeywords={Extended Reality, Intraprocedural Guidance, More},
}

\begin{document}
\maketitle

\begin{abstract}
Percutaneous epicardial access (PEA), performed on a beating heart under fluoroscopy, enables arrhythmia treatment. However, advancing a needle toward the thin, moving pericardium remains highly challenging and risky.
To address this challenge, we present a novel physics-driven sonification method for Extended Reality (XR)-based multisensory navigation to enhance user perception regarding critically sensitive needle landing phase in PEA.
Dynamic cardiac anatomy from 4DCTA was reconstructed and registered to a real-world coordinate system. 
Real-time needle tracking provided data of the needle tip relative to moving cardiac structures, driving the audio-visual feedback module.
The visual display presents navigational cues and dynamic cardiac structures, whereas the auditory display encodes physiological cardiac states into a rigorous mapping through a multilayered physical membrane.
To prove this novel concept, a phantom study was conducted with twelve cardiologists performing needle insertions under visual-only and multisensory feedback. 
Results show that the proposed multisensory method improved the navigation safety ($\chi^2 = 11.30, p < 0.01$), reducing myocardial contact (3.64\% vs. 7.27\%) and increasing correct access (90.91\% vs. 52.73\%). Needle placement accuracy improved, with closer membrane proximity (Cliff’s $\delta = 0.19$) and reduced variability (Fligner $p < 0.05$). Execution time remained comparable, but time–accuracy correlations differed significantly between modalities (Fisher’s $p < 0.01$). NASA-TLX showed lower cognitive load with multisensory guidance $(t = -4.02, p < 0.01)$. In conclusion, the multisensory system is associated with improved spatiotemporal awareness and suggests potential benefits for surgical navigation, while indicating reduced cognitive load. These findings demonstrate the feasibility of physics-driven sonification for supporting user-centered interaction and motivate further validation in realistic and clinical environments.
\end{abstract}


\section{Introduction}
\label{sec:introduction}

Percutaneous Epicardial Access (PEA) has emerged as a critical technique in electrophysiology for the treatment of a wide range of cardiac arrhythmias through catheter ablation. In PEA, a rigid needle is inserted near the sternum, just below the xiphoid, and advanced until it enters pericardial sac, i.e., a very thin compartment filled with fluid, by puncturing its outer membranous wall, which is called parietal pericardium. In its trajectory, the needle must not cross delicate tissues, such as diaphragm or lungs, and it must be stopped prior to puncturing the inner membranous wall of the pericardial sac, which is in continuity with the heart myocardium.
The adoption of PEA has rapidly increased, as demonstrated by a large single-center study~\cite{killu2016trends}, in which its use for ventricular arrhythmia treatment rose from 6 procedures in 2007 to 27–30 annually after 2010. This tendency is confirmed by a recent report indicating the utilization of PEA in the 23–30\% of ventricular tachycardia ablations~\cite{arya2025epicardial}.
Despite this broader adoption, PEA remains technically demanding. Because the needle is rigid, its insertion site and orientation determine the subsequent trajectory, yet these are selected solely by palpating bony anatomical landmarks~\cite{d2020percutaneous}.
During the advancement, the needle is visualized only on X-ray fluoroscopy images, but these provide only a two-dimensional (2D) projection of the needle and, upon the injection of contrast agent, of the pericardial silhouette, without yielding information on the surrounding tissues. Finally, given the limited information provided by X-ray fluoroscopy, it is very hard to stop the needle tip in the pericardial sac without perforating the myocardium. This challenge is further exacerbated by myocardial and diaphragmatic motion. 
Thus, accurately determining the final needle tip position is clinically relevant to verify entry into the pericardial space while avoiding myocardial perforation or misplacement before the insertion of subsequent catheters and guidewires for epicardial mapping~\cite{romero2019mastering}, and this process heavily depends on the operator’s experience, dexterity, and ability to interpret subtle differences in tissue resistance during needle advancement~\cite{boyle2012epicardial}.
As a matter of fact, a 6\% to 15\% risk of major PEA-related complications is reported~\cite{raad2024epicardial,mathew2021epicardial}. In a cohort of 271 epicardial ablation procedures, 34 major complications (12.5\%) were observed, including 23 cases (8.5\%) directly related to epicardial puncture~\cite{mathew2021epicardial}. The most common severe complications are pericardial bleeding resulting from inadvertent laceration of right ventricle and coronary vascular injury, with unintentional right ventricle puncture being reported in 4.5\% - 17\% of attempted epicardial ablation procedures~\cite{raad2024epicardial,romero2021epicardial}. Other severe complications, e.g., perforation of abdominal organs or collateral damage to epigastric arteries, have also been documented~\cite{raad2024epicardial,mathew2021epicardial}. 

Research in computer-aided surgery aimed to address these limitations and mitigate the risk of complications. Pre-procedural Electrocardiogram (ECG)-gated Time-resolved Computed Tomography Angiography (4DCTA) enables the reconstruction of a patient’s dynamic three-dimensional anatomy~\cite{kwong2015f}. By capturing Computed Tomography (CT) frames over at least four cardiac cycles and assuming minimal heart rate variability, this approach provides a reliable representation of both cardiac and respiratory motion. The resulting dynamic model supports precise preoperative planning, facilitating the identification of safe needle trajectories. Intraoperatively, ECG signal acquisition allows for temporal synchronization of the reconstructed dynamic models, while advanced registration methods~\cite{andrews2020registration,azampour2024anatomy,miura2025point} ensure their spatial alignment with the patient’s anatomy. Additionally, real-time tool tracking~\cite{ourak2021fusion,ramadani2022survey} and its integration within the 4DCTA coordinate system enable continuous monitoring of the needle’s position relative to the patient’s dynamic anatomy. 

Yet, a key challenge persists: effectively conveying this dynamic information to the operator during intraoperative navigation. Specifically, how can 4DCTA-based anatomical dynamic reconstructions such as the dynamic cardiac structures along with static planned trajectory, including the entry point and inclination of the needle, be presented directly onto the patient as navigational aids?

Recent advances in Medical Augmented Reality Systems (MARS) have demonstrated clear benefits for intraoperative guidance, enhancing situational awareness in neurosurgical, orthopedic, and oral procedures ~\cite{malhotra2023augmented}. However, these systems predominantly rely on unimodal visual displays, which are inherently constrained by shadows, occlusion, and limited depth perception, factors that compromise the accuracy of needle alignment and the targeting of static landmarks~\cite{heinrich2020comparison,heinrich20222d}. Furthermore, Optical See-through Head-Mounted Display (OST-HMD) devices suffer from limitations in stability, repeatability, and accuracy. These limitations are reflected in the display and user perception, leading to spatial misalignments and reduced reliability of visual overlays~\cite{soares2021accuracy,matyash2021accuracy,stadnytskyi2024experimental}.

These limitations pose significant challenges for developing effective surgical navigation systems that rely solely on unimodal visual cues~\cite{carbone2020commercially,condino2019perceptual}, particularly in procedures such as PEA. Recent studies~\cite{schutz2024framework,krumb2024cryotrack} have demonstrated the effectiveness of multisensory feedback by integrating auditory with visual cues. Specifically, auditory augmentation has been shown to enhance segmentation accuracy of anatomical structures in virtual reality setups~\cite{schutz2024framework} and to reduce execution time in abdominal cryoablation tasks~\cite{krumb2024cryotrack}.

Sonification, the process of conveying information through sound, has been shown to enhance user perception in interventional tasks involving static target structures, supporting situational awareness \cite{matinfar2018surgical}, surgical tool alignment \cite{black2017survey,matinfar2023sonification}, and the analysis of anatomical structures \cite{matinfar2024ocular}. 

Sound can convey multidimensional data over time through frequency, amplitude, and spectral patterns, all of which can be encoded with high resolution. The auditory system, with its remarkable temporal and spectral sensitivity, is well suited to decipher this complexity, perceiving pitch, rhythm, and timbre with precision. Its omnidirectionality enables perception independent of head orientation, thereby supporting continuous real-time interaction. Consequently, sonification allows for the rapid transmission of complex information, enhances engagement with computer-assisted surgery systems, and effectively preserves temporal continuity. These advantages have increasingly attracted research interest, particularly in presenting soft tissue deformations from tool interactions through auditory displays. For example, previous work~\cite{ruozzi2025biosonix} explores this potential in an offline setting, where sounds can be pre-recorded.

So far no methodologies exists allowing real-time accurate guidance in highly dynamic settings like PEA.

\textbf{Contribution} In this paper we present a multisensory feedback method designed to enhance needle navigation in fluoroscopy-guided PEA. The system was developed in close collaboration with medical experts to align with clinical requirements and is intended to complement rather than replace fluoroscopy. It provides real-time contextual awareness of the surgical instrument in relation to surrounding anatomical structures, with particular benefit as the needle approaches the pericardial membrane. Beyond this stage, conventional fluoroscopic views remain indispensable for final needle positioning and subsequent tool deployment within the epicardial space.

The primary objective of the proposed system is to provide precise guidance, ensuring that the needle reaches the pericardial membrane without penetrating the myocardium. While the overall workflow comprises multiple components, the focus of this study is on the evaluation of the intraoperative multisensory guidance. The complete workflow is described to provide the necessary context and to enable the accurate implementation and assessment of this guidance strategy. In this framework, an OST-HMD is used to display 4DCTA-based dynamic anatomy, static planned navigational cues, and a real-time needle avatar to establish spatial context, while sonification delivers detailed, dynamic guidance during navigation toward the beating heart.
To evaluate this approach, we conducted a phantom study with 12 cardiologists performing multiple insertions. Each participant completed the procedure twice, once using only visual guidance and once with the addition of sonification, allowing us to assess the impact of acoustic feedback on precision and control. 

The main contribution of this work is to address the challenge of needle guidance in the presence of dynamic targets. We propose and evaluate a novel Extended Reality (XR)-based multisensory navigation system designed to support situational awareness during the critical needle landing phase on beating cardiac structures in PEA. The system integrates visual and physics-driven auditory feedback, and its performance is assessed in a controlled experimental setting.

\section{Methods}
\label{sec:method}

\begin{figure}
    \centering
    \includegraphics[width=1.0\linewidth]{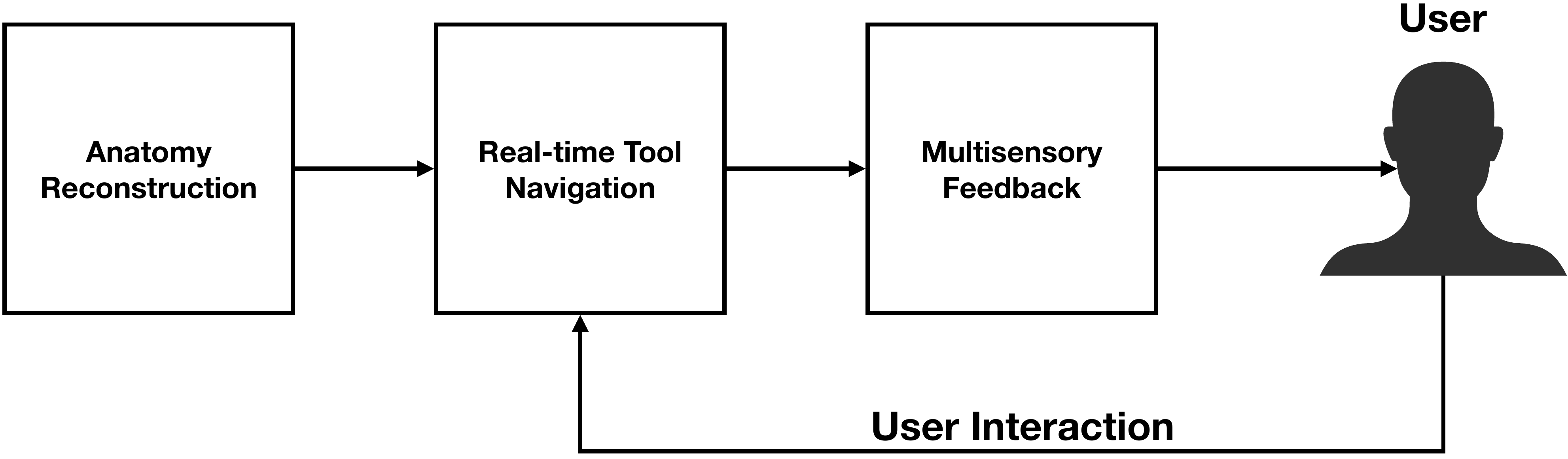}
    \caption{High-level schematic of the proposed workflow.}
    \label{fig:general-workflow}
\end{figure}

The proposed method is illustrated in Fig.~\ref{fig:general-workflow} as a modular workflow integrating three main components: (1) patient-specific anatomical reconstruction and surgical planning; 
(2) real-time tool navigation and the computation of navigation parameters serving as input to (3); 
and (3) multisensory feedback generation, with particular emphasis on physics-driven auditory cues, delivered to the user through an XR system comprising two display modalities: a visual interface implemented via an OST-HMD and an acoustic one provided through external speakers. 
These components operate in a closed-loop manner to ensure temporal and spatial coherence between tool motion and user guidance. 

\begin{figure}[htbp]
    \centering
    \includegraphics[width=1.0\textwidth]{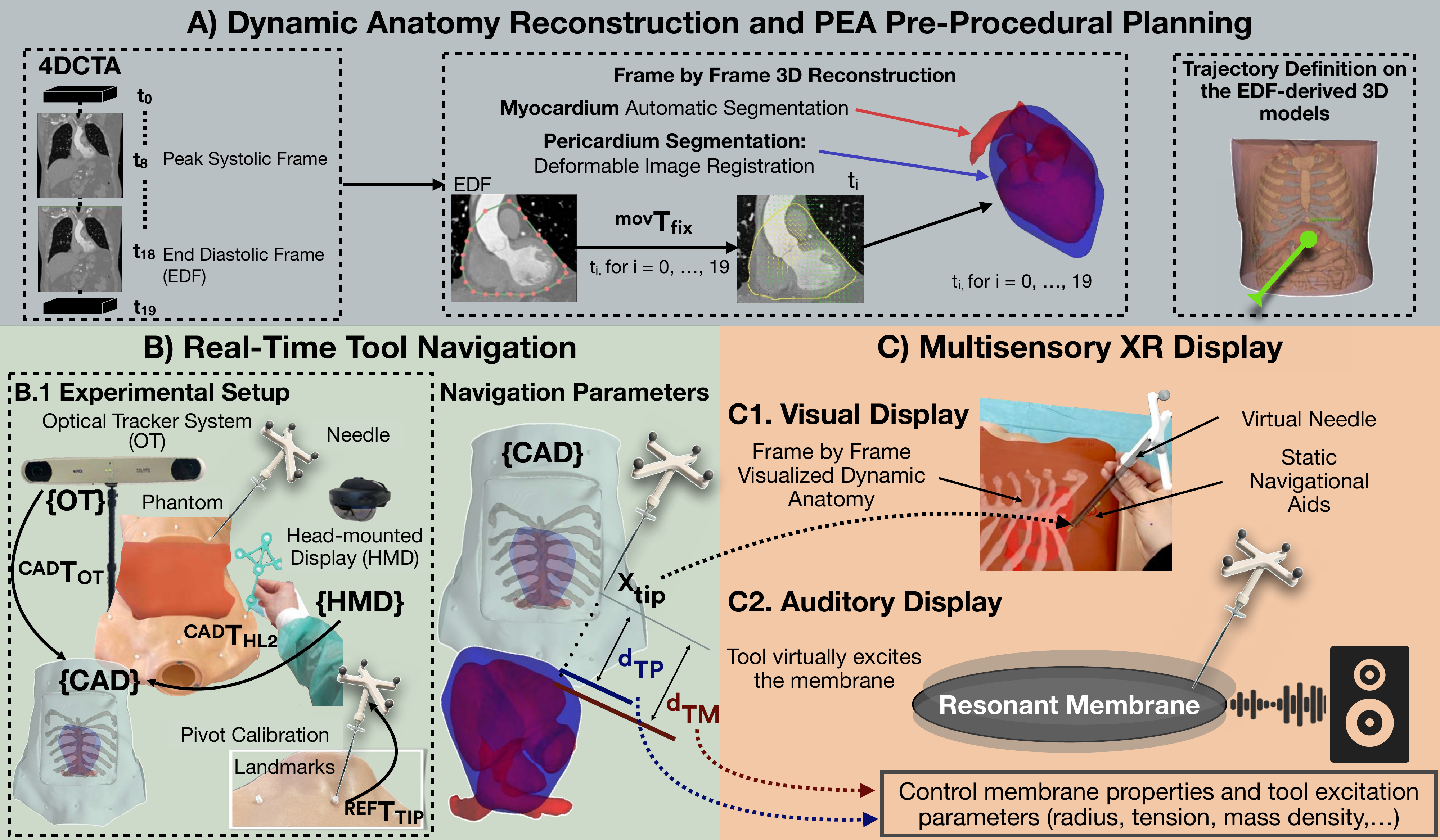}
    \caption{\textbf{Overview of the XR-based multisensory surgical navigation system developed for PEA.}
    \textbf{A)} ECG-gated 4DCTA frames over one cardiac cycle were analyzed. The myocardium and the external pericardial membrane (i.e. pericardium) were segmented, and their three-dimensional models reconstructed. An \textit{ad-hoc} planning algorithm was then applied to compute risk-optimized needle trajectories, avoiding surrounding critical organs. 
    \textbf{B)} Real-time computation of navigation parameters, defined as the Euclidean distance between the needle tip ($x_{tip}$) and the moving pericardium (\(d_{TP}\)) and myocardium (\(d_{TM}\)). In this study, a navigation setup for real-time needle tracking was implemented. Registration steps were required to coherently track the needle tip ($x_{tip}$) relative to the CT-derived phantom model (\{CAD\}), enabling the computation of navigation parameters. 
    \textbf{C)} The XR-based multisensory display integrated modular visual and auditory components with input data prepared as detailed in (B), to provide accurate and dynamic tool guidance.
 }
    \label{fig:pipeline}
\end{figure}

The detailed implementation of each component for PEA is shown in Fig.~\ref{fig:pipeline} and described in the following subsections (Sec.~\ref{sect::DynamicAnatomyReconstruction}, ~\ref{sect:toolnavigation}, ~\ref{sect:xrguidance}). However, the core contribution of this study lies in the methodology and evaluation of the multisensory guidance system (Fig.~\ref{fig:pipeline}C), implemented within an experimental setup for procedural simulation. Accordingly, the final subsection (Sec.~\ref{sect:evaluation}) focuses on the description of the experimental set-up and on the design and execution of the user study conducted to assess this component.

\subsection{Dynamic Anatomy Reconstruction and Trajectory Planning for PEA}
\label{sect::DynamicAnatomyReconstruction}

A preoperative ECG-gated 4DCTA dataset ($0.785 \times 0.785 \times 0.600 \,\text{mm}$ spatial resolution, 20 frames per cardiac cycle at 5\% temporal intervals) was analyzed (Fig.~\ref{fig:pipeline}A). 

Image processing was conducted in 3D Slicer~\cite{fedorov20123d}. The cardiac muscle (i.e., myocardium) was automatically segmented in each frame~\cite{wasserthal2023totalsegmentator}. The external pericardial membrane (i.e., pericardium), which presents greater segmentation complexity, was manually segmented in the End-Diastolic Frame (EDF). Segmentations for the other frames were obtained through deformable image registration~\cite{klein2009elastix}, by registering each frame to the EDF used as the fixed reference.
The resulting sequence of \(^{mov}T_{fix}\) transforms was applied to the pericardium segmentation mask, enabling dynamic pericardial profile segmentation across all time frames. The surfaces of the myocardium and the pericardium were reconstructed from the segmentation masks, and their distance (i.e. inter-parietal distance) was computed for each of the 20 frames. 
By analyzing the temporal variation of this distance throughout the cardiac cycle, safer needle landing zones, corresponding to regions of maximum inter-parietal distance, were identified. This information, combined with segmentation of critical organs (e.g., coronary, epigastric and mammary arteries, lungs, and liver) in the EDF frame, was used to initialize the pre-procedural planning platform.
Planning was performed considering the EDF as the worst-case scenario, since at this point in the cardiac cycle the myocardium is expected to reach its maximum volume, resulting in the minimum inter-parietal distance.
An \textit{ad hoc} algorithm was implemented to compute optimal trajectories through three consecutive filters: (1) the \textit{collision filter}: a ray-casting algorithm discarded any trajectory intersecting critical organs; (2) the \textit{length filter}: the total path length was compared against the needle length; and (3) the \textit{distance filter}: trajectories were refined to ensure a minimum clearance of 5 mm from all segmented sensitive structures, in accordance with clinical recommendations for epicardial punctures~\cite{aryana2020percutaneous,koruth2011unusual}.

\subsection{Real-Time Tool Navigation}
\label{sect:toolnavigation}
The proposed physics-driven sonification method (Sec.~\ref{sect:sonification}) relies on real-time needle tip tracking and the computation of navigation parameters. Specifically, the system continuously computes the Euclidean distances along the needle axis between the needle tip ($\mathbf{x}_{tip}$) and the moving pericardium and myocardium, denoted as $d_{TP}$ and $d_{TM}$, respectively (Fig.~\ref{fig:pipeline}B).

\subsection{Multisensory Extended Reality Surgical Navigation Display}
\label{sect:xrguidance}

The proposed XR surgical navigation system integrates complementary visual and auditory feedback modules to support precise tool guidance during pericardial access procedures. 
Both modules operate in real-time using the same input data stream, ensuring spatial and temporal coherence while maintaining full independence from each other. 

In the following sections, we describe the visualization and sonification modules and their respective data-processing pipelines.

\subsubsection{Visualization}
\label{sect:visualization}

The XR visual interface was built in Unity3D (Unity Technologies, v2020.3.1 LTS) and deployed to the OST-HMD device.
The 3D reconstruction of the patient-specific anatomy for 20 distinct time frames sampled over one full cardiac cycle (Sec.~\ref{sect::DynamicAnatomyReconstruction}) was visualized.
These reconstructed frames sequentially animated in a continuous loop during runtime, providing a dynamic reliable replica of the beating cardiac structures.
Static visual cues indicated the planned trajectory: a crosshair marks the access point, while a fixed cylinder aids needle alignment. An array of rings further enhances alignment\cite{chan2013visualization}. 
A virtual model of the needle was also added to the scene. Its pose was updated by the real-time $\textbf{x}_{tip}$, received via TCP.

\subsubsection{Sonification}
\label{sect:sonification}

For sonification to be reliable and effective, it must fulfill two key criteria: (a) \textbf{Informational Clarity:} The auditory representation must convey critical insights necessary for the safe execution of the task in a clearly perceptible manner. (b) \textbf{Intuitiveness and Usability:} The sonification must be easily comprehensible, allowing users to quickly learn and apply it in high-stakes surgical contexts. It should impose minimal cognitive load, ensuring that it does not interfere with the primary task.

\paragraph{Informational Clarity -- \textbf{State-Based Sonification}} To ensure safe puncture execution and accurate needle positioning within the epicardial space, the auditory representation must clearly convey critical spatial information relevant to the task. In particular, it should precisely indicate the tool-tip’s proximity to both the pericardial membrane and the myocardium. Furthermore, it must signal membrane perforation, enabling the user to halt needle advancement and achieve correct positioning without contacting the myocardium.

\begin{figure}
    \centering
    \includegraphics[width=0.8\linewidth]{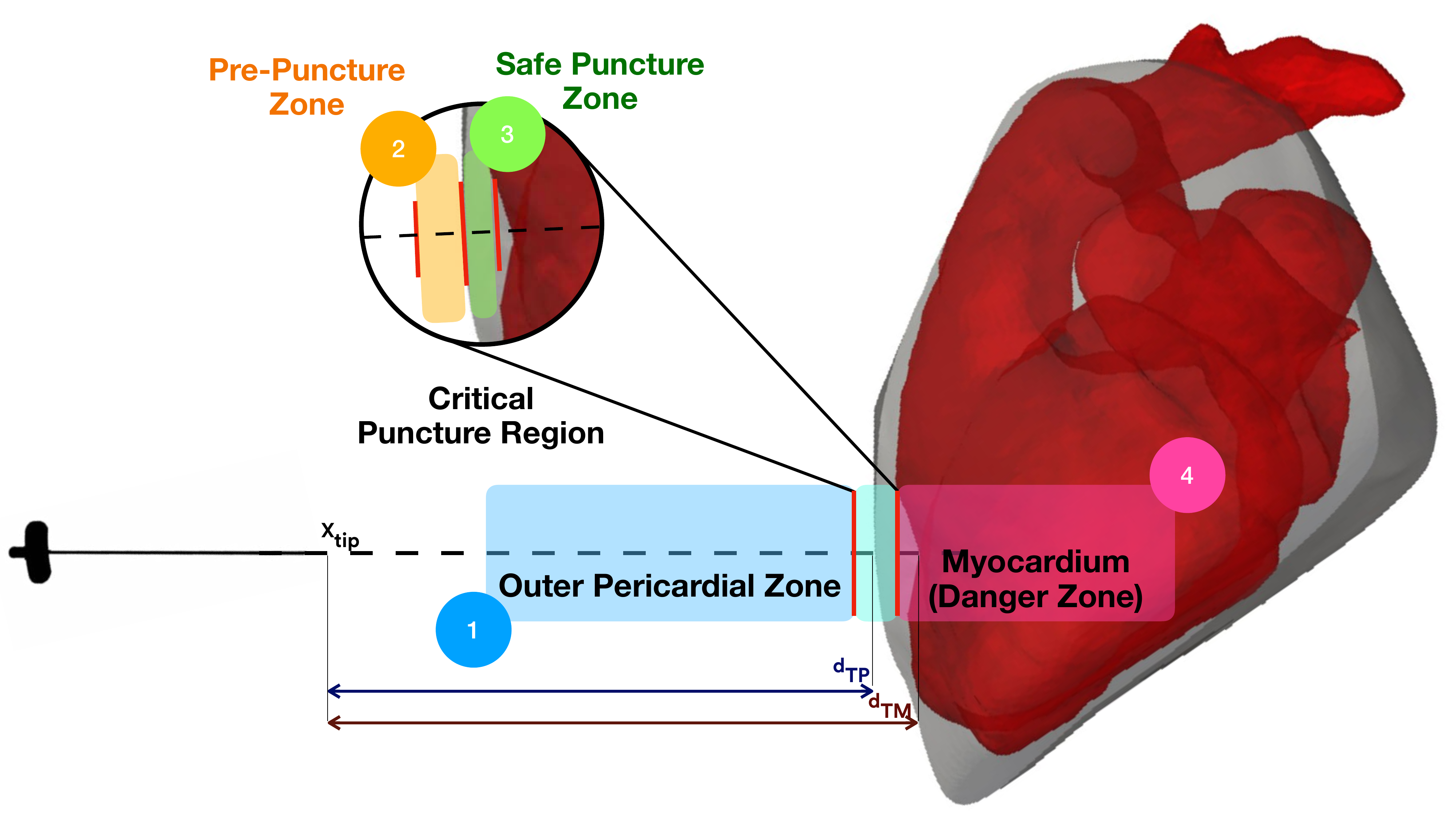}
    \caption{Conceptual representation of anatomical regions in PEA for state-based sonification. Transitions are defined based on the Euclidean distance computed along the needle axis from $\textbf{x}_{tip}$ to both the moving pericardium ($d_{TP}$) and myocardium ($d_{TM}$). The figure is for illustration purposes only; size and order of magnitude are not to scale.}
    \label{fig:four-states}
\end{figure}

To achieve this, we define four distinct \textbf{sonification states}, presented in Fig.~\ref{fig:four-states}, that must be clearly conveyed: \textit{(1) Outer Pericardial Zone,} the region outside the pericardium, where the user requires guidance to navigate toward the critical puncture area, encompassing key procedural stages. \textit{(2) Pre-Puncture Zone,} the area immediately adjacent to the pericardial membrane, indicating that the user is nearing, but has not yet reached, the puncture site. \textit{(3) Safe Puncture Zone,} the target region where the pericardial membrane has been successfully reached, and the puncture is safely performed. \textit{(4) Myocardium (Danger Zone),} a high-risk area that must be strictly avoided to prevent complications.

\paragraph{Intuitiveness and Usability -- \textbf{Physical Modeling Synthesis}} The complexity of this sonification scenario, involving multivariate and multistate signals, poses a challenge in delivering auditory cues that are both clearly conveyed and easily learnable, while ensuring they do not interfere with task execution. 

A straightforward one-to-one mapping approach, in which each data dimension is directly assigned to a single auditory parameter (e.g., mapping distance X to sound frequency), is simple to design but not necessarily optimal from an interaction perspective~\cite{Hermann2011PMSon}.

Conversely, a one-to-many mapping approach, in which each data dimension is represented by multiple auditory features, can enhance clarity and perceptual salience~\cite{Hermann2011PMSon}. However, improving usability requires sonification to provide causally driven sounds—namely, behaviorally familiar and physically meaningful auditory cues that enhance intuitiveness~\cite{theunissen2014neural,moerel2012processing}. Recent studies have demonstrated the effectiveness of embedding causal relationships in medical sonification~\cite{matinfar2023tissue,schutz2024framework}.

To address the challenges posed by heart dynamics and dual-layer distance, we employ physical modeling synthesis, which enables the definition of virtual objects that emulate realistic acoustic behaviors~\cite{smith1992physical,cook1997physically}. 

\paragraph{\textbf{2D Circular Membrane Model for Sonification}} We employed a 2D circular membrane model to represent the dynamic behavior of the heart structures. A circular membrane is a thin, flexible surface that vibrates when excited, similar to a drumhead. Its oscillatory behavior is governed by well-defined vibrational modes, which simultaneously determine multiple auditory parameters such as frequencies, amplitudes, and damping characteristics, thereby inherently realizing a one-to-many mapping. The membrane’s response is influenced by its radius, material properties such as surface density, tension, and damping factors, which control how energy dissipates over time. In this work, we used a subset of these factors for mapping, with Sec.\ref{sec:data_mapping} and Table\ref{tab:mappings} providing detailed parameterization of these factors. The membrane's oscillations combine to produce an output signal $s(t)$ derived from membrane vibrations that can be described as a sum of its vibrational modes:

\begin{equation}
    s(t) = \sum_{m,n} A_{m,n} e^{-\sigma_{m,n} t} \cos(2\pi f_{m,n} t + \phi_{m,n}),
\end{equation}

where $A_{m,n}$ is the amplitude of mode $(m,n)$, determined by the applied force; $f_{m,n}$ are the natural frequencies determined by the membrane’s physical parameters, including its radius, tension, and mass density, and $(m, n)$ are the mode indices, defining spatial vibration patterns; $e^{-\sigma_{m,n} t}$ represents damping, and is defined as: $e^{-(\beta + \alpha \cdot f_{m,n}^n) \cdot t}$, where  $\beta$  and  $\alpha$  are constant and frequency-dependent loss coefficients;  $\phi_{m,n}$ is the initial phase of each mode. 

For reproducibility, only the synthesis parameters explicitly defined in Section~\ref{sec:data_mapping} and Table~\ref{tab:mappings} were modulated during runtime, while all other membrane parameters were kept constant. The excitation force $F$ was held constant across all states, and the initial phase terms \(\phi_{m,n}\) were fixed at implementation defaults.

This formulation allows for realistic temporal decay of each mode, enhancing the acoustic representation of cardiac structures behavior.

\paragraph{\textbf{Data Mapping to Synthesis Parameters}} 
\label{sec:data_mapping}
The previously defined sonification states can be formally classified based on two input features, $d_{TP}$ and $d_{TM}$ using two threshold conditions, as follow:
\begin{itemize}
    \item State 1: $(d_{TP} > 5.0 \text{ mm}) \land (d_{TM} > 2.0 \text{ mm})$
    \item State 2: $(d_{TP} \leq 5.0 \text{ mm}) \land (d_{TM} > 2.0 \text{ mm})$
    \item State 3: $(d_{TP} \leq 0.0 \text{ mm}) \land (d_{TM} > 2.0 \text{ mm})$
    \item State 4: $(d_{TM} \leq 2.0 \text{ mm})$
\end{itemize}

State classification was performed using the original (non-normalized) distance values \(d_{TP}\) and \(d_{TM}\).

Each sonification state modulates one or more of the membrane’s parameters, influencing its dynamic response and resulting sound characteristics. The affected parameters are: $R$ (membrane radius, affecting both fundamental and modal frequencies; larger $R \Rightarrow$ lower $f_{m,n}$, thus lowering perceived pitch), $\beta$ (constant loss coefficient, controlling global damping; higher $\beta \Rightarrow$ shorter sustain via $\sigma_{m,n}$), $\alpha$ (frequency-dependent loss; higher $\alpha \Rightarrow$ stronger high-frequency damping, shaping spectral decay), $F$ (excitation force, determining mode amplitudes $A_{m,n}$), and $\Delta t$ (timing of force application, shaping rhythmic structure). 
This mapping dynamically modulates the membrane’s acoustic properties—pitch, damping, spectral content, and rhythmic excitation—based on input variations. 

The implementation modulates only the subset of parameters explicitly specified in Table~\ref{tab:mappings}. In practice, membrane radius $R$ was operationalized through the corresponding fundamental frequency $f_0$, which is reported in the table. The excitation force $F$ was kept constant across all states, while $\beta$, $\alpha$, and $\Delta t$ were dynamically updated according to the current state and normalized input values. Parameters not explicitly updated in a given state retained their previously assigned values within that state, ensuring continuity across state transitions.

The two distances ($d_{TP}$ and $d_{TM}$) were received via OSC in millimeters and normalized using the same linear clamping and scaling function. Each distance \(d \in \{d_{TP}, d_{TM}\}\) was mapped to a normalized variable \(\hat{d} \in [0,1]\) as
\[
\hat{d} = \frac{\mathrm{clip}(d, d^{\min}, d^{\max}) - d^{\min}}{d^{\max} - d^{\min}},
\]
where \(\mathrm{clip}(x,a,b)=\min(\max(x,a),b)\). The bounds were set to \(d_{TP} \in [1,60]\) mm and \(d_{TM} \in [1,30]\) mm.

The normalized variables \(\hat{d}_{TP}\) and \(\hat{d}_{TM}\) directly determine the parameter ranges and interpolation intervals specified in Table~\ref{tab:mappings}.

State assignment was evaluated hierarchically in the following priority order: State 4 \(\rightarrow\) State 3 \(\rightarrow\) State 2 \(\rightarrow\) State 1, to avoid ambiguity when threshold conditions overlap. Within each state, parameter values were updated by linear interpolation over the corresponding normalized input intervals specified in Table~\ref{tab:mappings}.

For consistency with Table~\ref{tab:mappings}, normalized variables are expressed in descending intervals from 1 (far) to 0 (contact).

\begin{table}[t]
\centering
\caption{\textbf{State-Dependent Parameter Modulation.} The normalized input variables \(\hat{d}_{TP}\) and \(\hat{d}_{TM}\) define the interpolation intervals within each state. Membrane pitch is reported as fundamental frequency \(f_0\), which operationalizes membrane radius \(R\). Temporal excitation intervals \(\Delta t\) are given in milliseconds.}
\label{tab:mappings}
\setlength{\tabcolsep}{3pt}
\renewcommand{\arraystretch}{1.3}
\scriptsize
\begin{tabular}{|c|c|c|c|c|}
\hline
\textbf{State} & $\mathbf{f_0}$(Hz) & $\mathbf{\beta}$ & $\mathbf{\alpha}$ & $\mathbf{\Delta t}$(ms) \\ \hline
1 & 100 & 2 & 10 & $d_{TP} \in [1,0.5)$ \\
  &     &   &    & $\rightarrow \Delta t \in [500,271]$ \\ \hline
2 & 100 & $d_{TM} \in [1,0.5)$ & $d_{TM} \in [1,0.5)$ & $d_{TP} \in [0.5,0)$ \\ 
  &     & $\rightarrow \beta \in [2,1.06]$ & $\rightarrow \alpha \in [10,5.075]$ & $\rightarrow \Delta t \in [270,40]$ \\ \hline
3 & 400 & $d_{TM} \in [0.5,0]$ & $d_{TM} \in [0.5,0]$ & 40 \\  
  &     & $\rightarrow \beta \in [1.05,0.1]$ & $\rightarrow \alpha \in [5.075,0.15]$ & \\ \hline
4 & 1000 & 0.1 & 0.15 & 40 \\ \hline
\end{tabular}
\end{table}

These mappings are designed to ensure that the resulting sounds fall not only within the range of optimal auditory perception but also within a perceptual comfort zone, where differences are clear, easily distinguishable, and not fatiguing to the listener~\cite{zwicker2013psychoacoustics}. The parameter values in Table~\ref{tab:mappings} were selected to create perceptually distinct sonic spaces. Fundamental frequencies of 100, 400, and 1000~Hz are separated by more than one auditory critical bandwidth ($\approx 100$~Hz at 400~Hz, $\approx 160$~Hz at 1000~Hz), ensuring robust pitch discrimination~\cite{zwicker2013psychoacoustics}. The damping parameters $\alpha$ (10--0.15) and $\beta$ (2--0.1) control the exponential decay of the oscillatory response, shaping loudness envelopes with differences comfortably above detection thresholds of $\sim 1$~dB and thereby yielding clearly distinct decay behaviors~\cite{moore2012introduction}. The temporal parameter $\Delta t$ spans 500--40~ms, covering perceptual tempo categories from slow ($>500$~ms) to fast ($<100$~ms), with differences well beyond the $\sim 20$--30~ms temporal resolution limit~\cite{zwicker2013psychoacoustics}. Altogether, this parametrization provides a systematic basis for mapping physical dynamics to perceptually effective auditory cues.

With this approach, perceptual sound attributes are not prescriptively selected, as in conventional parameter-mapping methods; instead, they emerge from the resonator dynamics, shaped jointly by the initial model parameters and their anatomy-inspired, state-dependent, and interaction-driven modulations.

\paragraph{\textbf{Cognitive Auditory Mapping and Learned Associations}} This mapping ensures that variations in input data dynamically modulate the membrane’s acoustic properties, influencing pitch, damping, spectral content, and rhythmic excitation patterns. The rhythmic excitation follows an echolocation-inspired mechanism, analogous to that used in parking assistance systems, serving as a fundamental cue for distance perception. Energy dissipation (regulated by $\alpha$ and $\beta$) emulates the myocardium’s dynamic damping behavior during diastole and systole, where varying proximity to the pericardial membrane modulates resonance and provides an intuitive, perceptually salient auditory cue. In addition, modulating the membrane radius ($R$) induces pitch variation, reinforcing perceptual distinctions between interaction zones. When the tool reaches the myocardium threshold, the radius, damping coefficients ($\alpha$, $\beta$), and spectral characteristics shift abruptly, producing a clear and immediate auditory warning of critical proximity. 

It is worth noting that the resulting auditory feedback emerges from a model that reflects anatomical context and user interaction, preserving causal coherence between interaction, anatomical state, and sound. This enables systematic and extensible sonification in complex, time-varying surgical scenarios such as needle guidance under cardiac motion. In contrast, conventional approaches predominantly rely on explicit parameter-mapping strategies that directly map low-dimensional signals to sound parameters (e.g., pitch or rate), often requiring manual signal segmentation and heuristic tuning as interaction complexity increases.

A supplementary video (S1) is provided to illustrate the auditory mapping framework and to demonstrate the resulting sounds.

\subsection{System Evaluation: The Phantom Study}
\label{sect:evaluation}

To evaluate the proposed XR-based multisensory surgical navigation system in a controlled yet clinically meaningful environment, we conducted a phantom study. This section describes the experimental setup used to simulate PEA and the design of the user study. 

\subsubsection{Experimental Setup}
\label{sect:exptoolnavigation}

The experimental setup (Fig.~\ref{fig:pipeline}B.1) consisted of: (i) a phantom with a rigid sternum, ribs, and a puncturable foam chest patch; (ii) an optical tracker (NDI Polaris Spectra); (iii) a 18 gauge Tuohy needle (150 mm long) with a custom three-dimensional (3D)-printed passive marker array; and (iv) an OST-HMD device. 
A CT scan of the phantom was acquired to generate its 3D virtual reconstruction. Since the physical phantom did not include cardiac structures, its reconstruction was added to the virtual scene described in Sec.~\ref{sect:visualization} and registered to the animated cardiac anatomy by aligning the sternum structures.

In this XR framework, three Frame of Reference (FoR) were involved: the optical tracker (i.e. \{OT\}), the phantom anatomy reconstructed from the CT scan (i.e. \{CAD\}), and the OST-HMD FoR (i.e. \{HMD\}). 
To enable real-time computation of the navigation parameters such as the needle tip's relative distance to the anatomy, and to ensure the coherence of the information among all the FoRs, calibration and registration procedures were required.
First, the transformation \(^{REF}T_{TIP}\) was computed to correctly track the needle's tip with respect to the optical tracker FoR (\{OT\}), achieving a pivot calibration Root Mean Square Error (RMSE) of 1.09 mm. 
Considering the \{CAD\} as the global FoR, two landmark-based registration procedures with the \{OT\} and the \{HMD\} were performed. Eight landmarks were designed to accommodate the needle's tip in a unique manner, and were attached to the phantom. Their precise location within the \{CAD\} were reconstructed from the CT scan.
Landmarks' positions were acquired both in the \{OT\} space with the tracked needle, and in the \{HMD\} space using another tool tracked directly by the OST-HMD~\cite{iqbal2022semi}. The accuracy of the landmark-based registration was evaluated by computing the RMSE. 
The transformation from \{OT\} to \{CAD\} space (i.e. \(^{CAD}T_{OT}\)) yielded an RMSE of 1.2 mm, whereas the registration of \{HMD\} with \{CAD\} space (i.e. \(^{CAD}T_{HMD}\)) resulted in an RMSE of 2.2 mm. 

This procedure ensured a correct tracking of the needle's tip in the \{CAD\} space (i.e. $x_{tip}$).
This information was streamed in real-time via the Transmission Control Protocol (TCP) to OST-HMD. 
Moreover, $d_{TP}$ and $d_{TM}$ were computed and streamed to the sonification module via the Open Sound Control (OSC) protocol~\cite{wright20172003}. 

\subsubsection{The User Study}
\label{sect:userstudy}
The user study was co-designed with medical experts to ensure clinical relevance and alignment with task requirements. The goal was to assess whether the proposed multisensory guidance improves safe needle navigation toward the pericardial membrane compared to visual-only guidance. Our system is intended to support the operator specifically when the needle approaches the pericardial membrane. Guidance success was defined as reaching the pericardial membrane while avoiding contact with the underlying beating myocardium.
Participants were asked to perform a series of punctures by following pre-defined trajectories. As the primary focus of this study was dynamic target navigation, the initial needle-to-path alignment was implemented following the method presented in~\cite{matinfar2023sonification} and was not part of the evaluation. Each trial started with a start command once correct alignment was established and ended when the participant verbally confirmed completion by saying \textit{“STOP”}, indicating user's confidence in the needle position at which they would have proceeded with the next procedural steps. 

To assess the impact of sonification, we conducted a user study comparing multisensory guidance and visual-only guidance during needle navigation to the pericardial membrane.

Relying the experimental navigation system (Sec.~\ref{sect:exptoolnavigation}), sonification was implemented via MAX/MSP\footnote{https://cycling74.com/products/max} for real-time audio processing and Modalys\footnote{https://support.ircam.fr/docs/Modalys/current/} for physical modeling synthesis (Sec.~\ref{sect:sonification}). 

\begin{figure}
    \centering
    \includegraphics[width=0.5\linewidth]{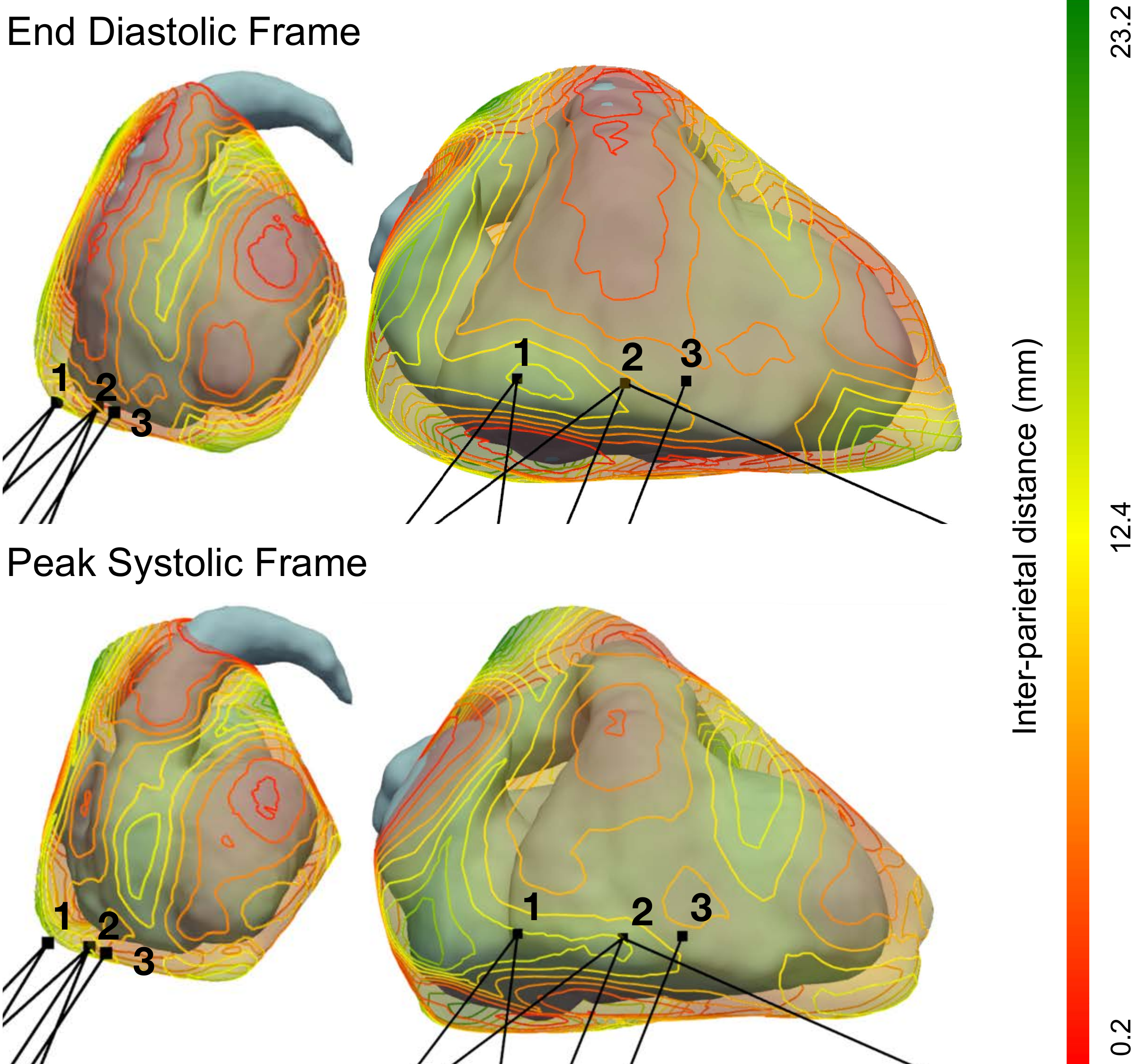}
    \caption{Planned trajectories for the phantom study, shown at End-Diastolic and Peak-Systolic phases of the cardiac cycle. Six trajectories were generated toward three target points (1,2,3) on the pericardium, following the strict requirements of PEA (Sec.~\ref{sect::DynamicAnatomyReconstruction}). The myocardium and the initial tract of the pulmonary artery are shown in gray. The pericardium is color-coded to represent the inter-parietal distance between the myocardium and pericardium. This distance varies both spatially and temporally: it increases during Peak Systole, whereas End Diastole represents the most critical phase, with the myocardium at maximum volume and a higher risk of myocardial injury.}
    \label{fig:plannedpath}
\end{figure}

Six trajectories were generated using the planning algorithm (Sec.~\ref{sect::DynamicAnatomyReconstruction}), targeting three distinct points (1, 2, 3) on the pericardial sac (Fig.~\ref{fig:plannedpath}). A medical expert validated the trajectories for consistent difficulty and clinical relevance, and they were presented in uniformly randomized order to the participants. The assignment of trajectories was balanced across modalities to ensure comparable task difficulty between V and MS conditions.

The target population for the study consisted of cardiologists and electrophysiologists. 
They had no prior experience with our system: they first received a brief introduction, followed by a standardized explanation of the study procedure ($\backsim5$ min). They then performed a self-exploration of the system, completing three punctures with visual feedback and three with multisensory feedback ($\backsim5$ min). This familiarization phase was designed to provide comparable initial exposure to both modalities and reduce potential learning effects during the evaluated trials.

Each participant performed ten trials: five punctures under visual-only (V) guidance and five under multisensory (MS) guidance. The modality order was counterbalanced across participants (V–MS vs. MS–V) to mitigate potential order and learning effects, with a short break of approximately 5 minutes between blocks to reduce carry-over effects. Within each modality block, trial order was randomized to further minimize sequence effects.

Execution time was recorded from the start command to the participant’s verbal confirmation of completion.
Throughout each puncture, the needle tip's intersections with the dynamically moving cardiac structures were continuously monitored with a ray casting–based collision detection algorithm. Specifically, potential intersections with the pericardium and myocardium were recorded during the whole task within the 20-frame cardiac cycle animation. 
Additionally, the final needle tip position at the moment of completion was logged.
After each condition, participants completed the National Aeronautics and Space Administration Task Load Index (NASA-TLX) workload questionnaire, rated perceived difficulty of each modality, and provided subjective feedback.

This experimental design, combining counterbalanced modality order, randomized trial sequences within each block, and prior familiarization, was designed to mitigate potential learning and carry-over effects between conditions.

\section{Results} 
\label{sect:results}

The experimental data collected as described in Sec.~\ref{sect:userstudy} were analyzed investigating several aspects relevant for intraoperative guidance in PEA.

First, navigation task was assessed by determining whether participants were able to safely reach the pericardium surface while avoiding myocardial contact.

Subsequently, the final needle tip position at the moment of the \textit{“STOP”} command was examined, considering only safe (successful) trials.
Since the system was designed to guide the operator in accurately reaching the pericardium, the needle tip position at the point when participants felt confident to stop was considered relevant, as it represents the starting point of subsequent procedural steps under fluoroscopic guidance (as outlined in the Introduction). At this stage, the needle was fixed within the phantom’s foam chest patch; thus, the recorded tip position reflected the operator’s perceived alignment with the intended target. 
In contrast, the final needle tip orientation was not evaluated because the study focused on dynamic target navigation rather than trajectory-following accuracy, with initial needle-to-path alignment handled as in~\cite{matinfar2023sonification}. Furthermore, orientation at the end of advancement has limited clinical relevance once safe pericardial entry is achieved, as it does not impose strict constraints on subsequent catheter deployment.
This final position was therefore used to assess the accuracy of needle placement according to two aspects: (1) accuracy of the pericardial access: once the sac was reached, how promptly did the user perceive the moving membrane and stop the needle, while maintaining minimal penetration depth to ensure tip placement suitable for subsequent catheter deployment? This was quantified as the minimum distance from the final needle tip position to the moving membrane; and (2) the spatial accuracy: where was the final needle tip located relative to the predefined anatomical target for each trajectory? This was quantified as the Euclidean distance between the final needle tip and the planned target point.

This section is structured reporting the results of these analysis.

The study included 12 participants (age range: 25--47 years; 10 male, 2 female), comprising 2 cardiologists, 3 electrophysiologists, and 7 medical residents with heterogeneous experience in PEA. 

The participants exhibited diverse levels of procedural experience: 3 performed PEA weekly, 2 had completed it fewer than five times, 3 had assisted in procedures, and 4 had no prior experience. Most reported limited musical practice (either passive listening or rare instrument use) and variable exposure to OST-HMD technology, ranging from no prior use to occasional experimentation. 

Supplementary Video (S2) provides a demonstration of the user study conducted in the experimental set-up.

\subsection{Navigation Task Outcome Analysis} 
Each trial was evaluated by classifying the navigation outcome into three categories based on the intersection data recorded during navigation up to the user’s verbal \textit{"STOP"} confirmation. Each participant performed ten trials (five under V guidance and five under MS guidance):

\begin{itemize}
\item \textit{Successful Completion}: reaching the pericardial membrane without myocardial contact;
\item \textit{Missed Target}: failure to reach the pericardial sac during the navigation;
\item \textit{Critical Failure}: unintended contact with the myocardium.
\end{itemize}

The study comprised 120 evaluated trials in total (12 participants × 10 trials), including 60 trials under V guidance and 60 under MS guidance. Task outcome rates were computed over all trials within each modality (n = 60 per condition). 

For both modalities, we computed the distribution of task outcome rates according to the three categories defined above and performed Chi-square tests to evaluate statistical significance between the groups. 

Across all participants, the \textit{Successful Completion} rate was higher under MS guidance (83.3\%, 50/60) than with V guidance (55.0\%, 33/60). Conversely, the incidence of \textit{Missed Targets} was lower under MS (13.3\%, 8/60) compared to V (36.7\%, 22/60). Similarly, the rate of \textit{Critical Failures} decreased with MS (3.3\%, 2/60) relative to V (8.3\%, 5/60). Statistical analysis confirmed a significant effect of modality on task outcome ($\chi^2 = 11.30, p < 0.01$).

\textbf{Expertise Level Subgroups:} An independent analysis examined performance differences between novices (n = 7) and experts (n = 5) under each modality.

\begin{itemize}
\item Under \textbf{visual-only} guidance, novices achieved a \textit{Successful Completion} rate of 54.29\%, compared to 56.00\% for experts. \textit{Missed Target} occurred in 37.14\% (novices) and 36.00\% (experts) of trials. \textit{Critical Failures} occurred in 8.57\% (novices) and 8.00\% (experts).
\item Under \textbf{multisensory guidance}, novices achieved 91.43\% \textit{Successful Completion} rate and experts 72.00\%. \textit{Missed Targets} were rare (8.57\% novices; 20.00\% experts). \textit{Critical Failures} were absent for novices (0.00\%) but occurred in 8.00\% of experts trials.
\end{itemize}

Within each modality, no statistically significant differences were found between novices and physicians (V: $\chi^2 = 0.37, p = 0.829$; MS: $\chi^2 = 5.13, p = 0.074$).

\begin{figure}[!t]
    \centering
    \includegraphics[width=0.5\linewidth]{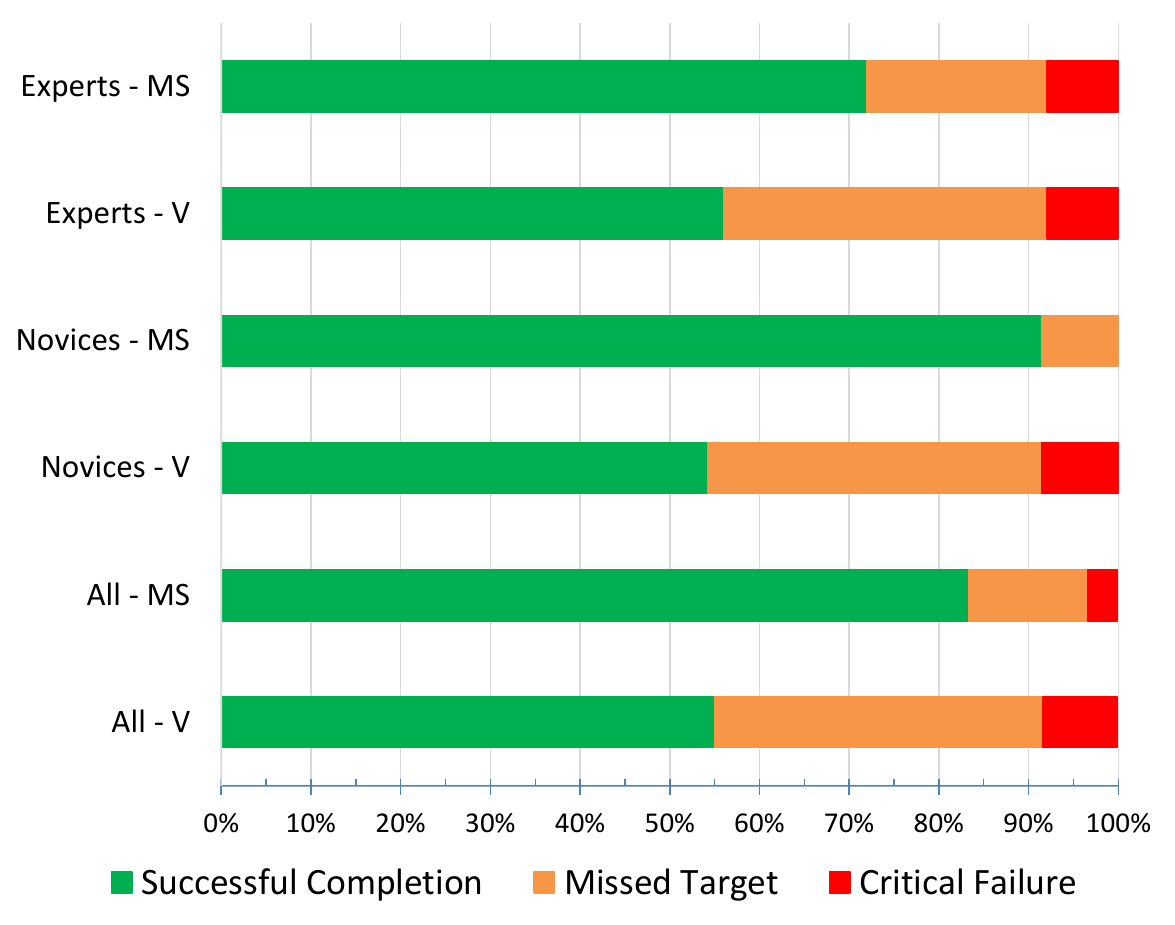}
    \caption{
        Distribution of task outcome rates (\textit{Successful Completion}, \textit{Missed Target}, and \textit{Critical Failure}) 
        under visual-only (V) and multisensory (MS) modalities for all participants, novices, and experts. 
        The stacked bars highlight the higher success rates and lower error rates under MS across all groups.
    }
    \label{fig:accuracy}
\end{figure}

The outcome rates across expertise groups and modalities are summarized in Table~\ref{tab:accuracy} and illustrated in Fig.~\ref{fig:accuracy} using a stacked bar plot.

\begin{table}[htbp]
    \centering
    \caption{
        Task outcome rates for expertise group under visual-only (V) and multisensory (MS) modalities. 
        All the enrolled participants (12) performed 5 tasks per modality. Experts (5) and novices (7) subgroups were also analyzed.
        The higher values by \textit{Success} and the lower values by \textit{Missed Target} and \textit{Critical Failure}, which show better performance, are bolded.
    }
    \renewcommand{\arraystretch}{2.0}
    \setlength{\tabcolsep}{4pt}
    \begin{tabular}{l|c|c|c|c}
        \textbf{Group} & \textbf{Modality} & \textbf{Success} & \textbf{Missed Target} & \textbf{Critical Failure} \\ \hline
        \multirow{2}{*}{\textbf{All} (n=12)} 
            & V  & 55.00\% & 36.66\% & 8.33\% \\
            & MS & \textbf{83.33}\% & \textbf{13.33}\% & \textbf{3.33}\% \\ \hline
        \multirow{2}{*}{\textbf{Novices} (n=7)} 
            & V  & 54.28\% & 37.14\% & 8.57\% \\
            & MS & \textbf{91.43}\% & \textbf{8.57}\% & \textbf{0.00}\% \\ \hline
        \multirow{2}{*}{\textbf{Experts} (n=5)} 
            & V  & 56.00\% & 36.00\% & 8.00\% \\
            & MS & \textbf{72.00}\% & \textbf{20.00}\% & 8.00\% \\
    \end{tabular}
    \label{tab:accuracy}
\end{table}

\subsection{Needle Tip Placement Analysis} 
\label{sect:tipaccuracy}
Analyses of final needle-tip placement were restricted to successful trials only, yielding 33 samples for V and 50 samples for MS.

\begin{figure}[!t]
    \centering
    \includegraphics[width=0.7\linewidth]{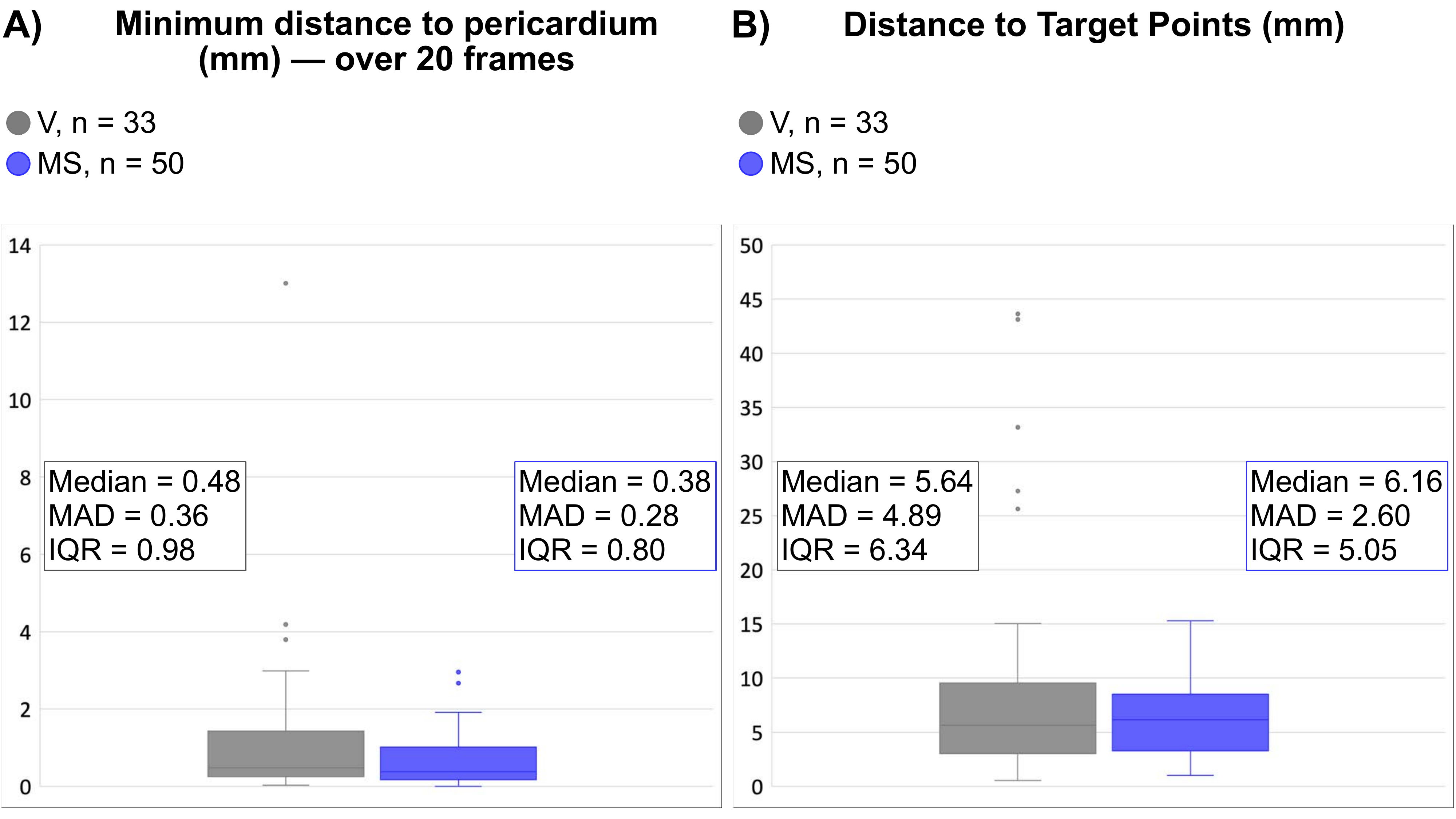}
    \caption{Box plot visualization of successful needle placement accuracy evaluated under two aspects : (A) stopping advancement upon reaching the pericardial membrane; and (B) accurately reaching the pre-planned target points (). The $n$ variable represents the number of samples considered.}
    \label{fig:box_plot}
\end{figure}

Table~\ref{tab:accuracy_tip} summarizes the key descriptive statistics for both analyses, and Fig.~\ref{fig:box_plot} schematically shows them. The variable $n$ represents the number of considered samples for each modality.
Given the non-normal distribution of the data, we report the median alongside measures of dispersion, Mean Absolute Deviation (MAD) and Interquantile Range (IQR), as well as the range (Min and Max).

\begin{table}[!t]
\renewcommand{\arraystretch}{1.4}
\centering
\caption{Summary of accuracy measurements for final needle tip placement. Values are reported for successful trials only. $n$ is the number of samples considered for each modality.}
\label{tab:accuracy_tip}
\begin{tabular}{lcccccc}
\toprule
\textbf{Modality} & \textbf{n} & \textbf{Median} & \textbf{MAD} & \textbf{IQR} & \textbf{Min} & \textbf{Max} \\
\addlinespace 
\midrule
\multicolumn{7}{c}{\textbf{Minimum Distance to Pericardium (mm) -- across 20 frames}} \\
V   & 33 & 0.48 & 0.36 & 0.98 & 0.03 & 13.01 \\
MS  & 50 & 0.38 & 0.28 & 0.80 & 0.00 & 2.95 \\
\addlinespace 
\midrule
\multicolumn{7}{c}{\textbf{Distance to Target Points (mm)}} \\
V   & 33 & 5.64 & 4.89 & 6.34 & 0.53 & 43.64 \\
MS  & 50 & 6.16 & 2.60 & 5.05 & 1.00 & 15.30 \\
\addlinespace 
\bottomrule
\end{tabular}
\end{table}

\subsubsection{Distance to the Pericardium}
\label{sect:precision_to_peri}

Needle placement accuracy relative to the pericardium surface was quantified as the minimum distance between the final tip position and the pericardium mesh across the 20-frame cardiac cycle. In this formulation, smaller values reflect higher accuracy, as they indicate that the user more promptly and reliably perceived reaching the membrane and stopped the needle, while inside the sac.

With visual-only guidance, the median distance to the pericardium was \SI{0.48}{\milli\meter} (IQR: \SI{0.98}{\milli\meter}, range: \SIrange{0.03}{13.01}{\milli\meter}), while under multisensory guidance it was lower at \SI{0.38}{\milli\meter} (IQR: \SI{0.80}{\milli\meter}, range: \SIrange{0.00}{2.95}{\milli\meter}). 

A non-parametric Mann–Whitney U test was chosen because the data were not normally distributed and group sizes were unequal.
A Mann-Whitney U test showed no statistically significant difference between modalities in distance to the pericardium ($U=982, p=0.145$), though a small effect size (Cliff's $\delta=0.19$) suggested a marginal advantage for multisensory guidance. Fligner test indicated no statistical significance in dispersion ($p=0.1385$).

\begin{figure}[!t]
    \centering
    \includegraphics[width=0.7\linewidth]{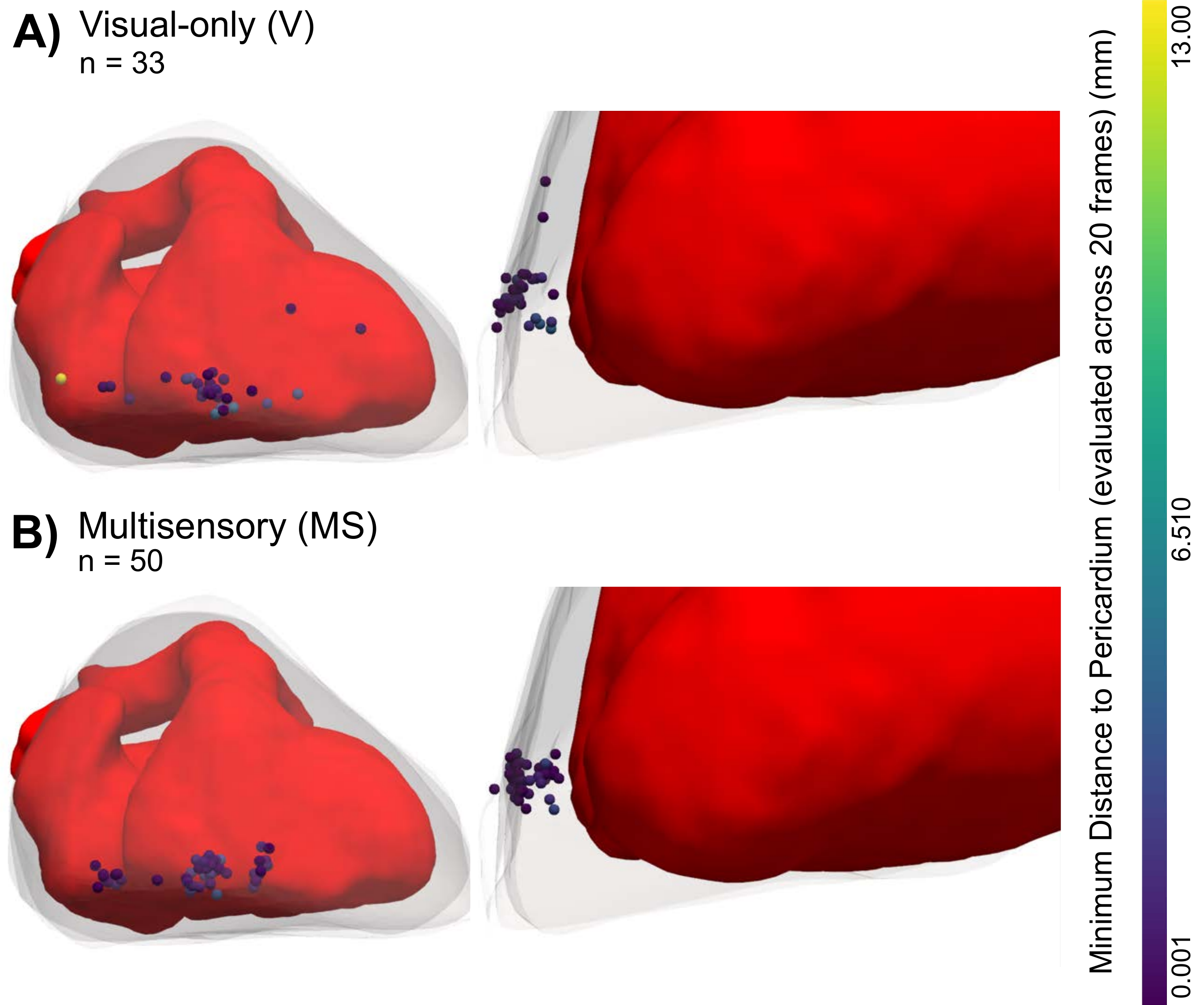}
    \caption{Final successful needle tip placements for both modalities relative to the pericardial membrane. Colors indicate the minimum distance to the dynamically moving pericardium (evaluated across 20 frames). The multisensory condition (bottom) exhibits greater spatial consistency, with placements more closely aligned to the membrane.}
    \label{fig:accuracy_to_peri}
\end{figure}

Fig.~\ref{fig:accuracy_to_peri} illustrates these distributions, indicating closer proximity to the pericardium under multisensory guidance.

\subsubsection{Distance to the Target Points}
\label{sect:precision_to_target}

\begin{figure*}[!t]
    \centering
    \includegraphics[width=0.7\linewidth]{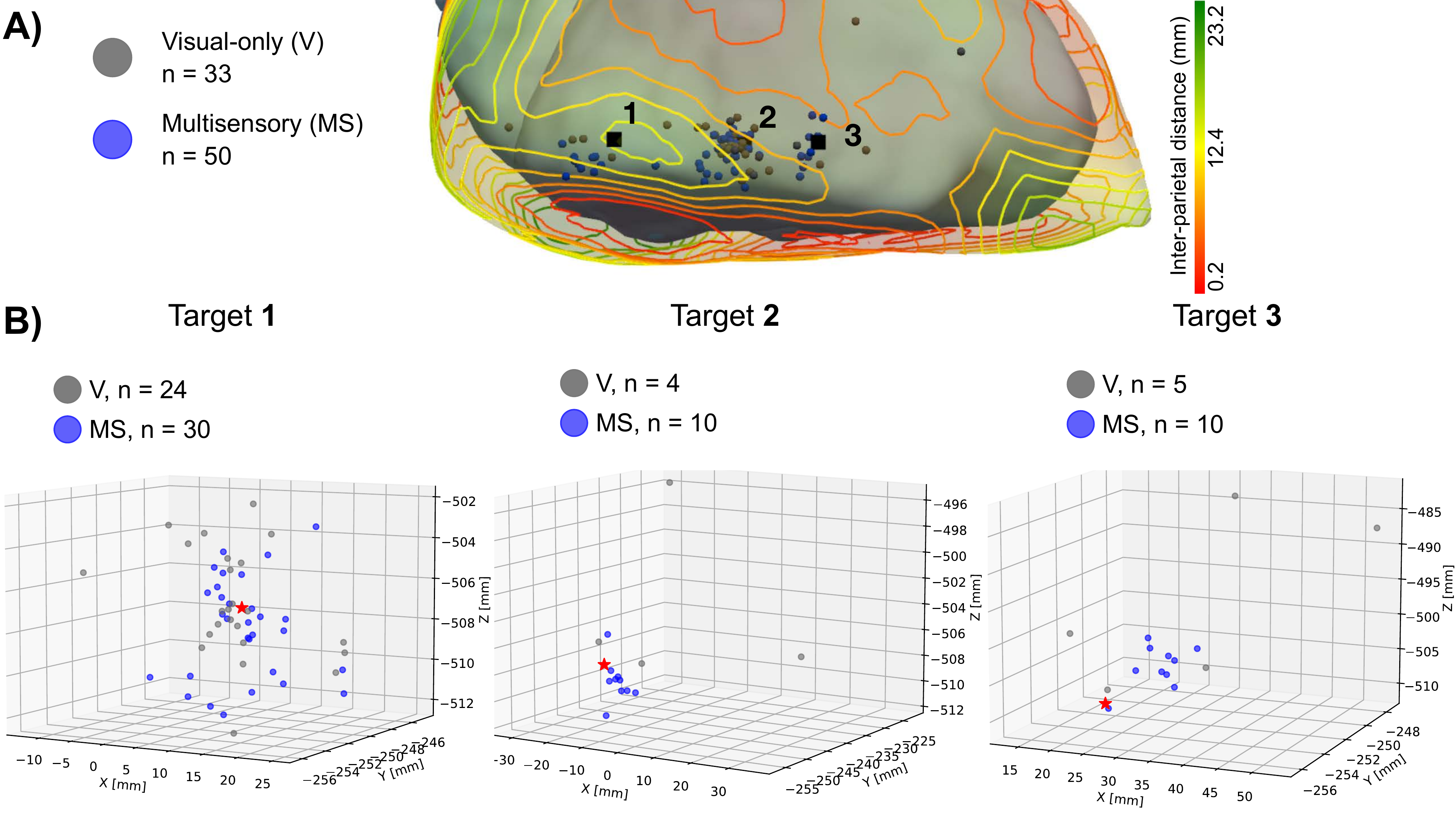}
    \caption{Needle tip placement relative to the planned target points 1, 2, 3 across both modalities for successful placements only. The value of considered samples for each test is reported ($n$).
    \textbf{(A)} 3D distribution of final tip positions relative to the pericardial surface (End-Diastolic phase). The clustering effect demonstrated by the MS modality, helps to reduce high-risk placements with tips avoiding critical areas of cardiac wall proximity.
    \textbf{(B)} Each subplot shows the spatial spread of final tips. While both modalities exhibit deviations from the planned targets, MS generally resulted in tighter clustering and reduced spread compared to V.}
    \label{fig:accuracy_to_target}
\end{figure*}

We further analyzed the Euclidean distance between the final needle tip position and the correspondent planned target points.
These target points are defined in the End Diastolic frame which represents the most critical situation, with the myocardium at maximum volume and higher risk of myocardial injury. 
These points satisfied all the pre-planning safety criteria (see Sec.~\ref{sect::DynamicAnatomyReconstruction}).
Thus, lower distance to these points, ensures safer needle placement and avoids high-risk punctures.
Fig.~\ref{fig:accuracy_to_target} illustrates the 3D distributions of needle tip placements across modalities and targets, along with the corresponding number of samples ($n$).
Statistical testing between the two modalities revealed no significant difference in medians (Mann–Whitney $U=847.0000$, $p=0.8414$) and only a negligible effect size ($\delta < 0.147$). In contrast, the Fligner test indicated a significant difference in dispersion ($p<0.05$).
  
\subsection{Execution Time} 

Execution times of the whole collected samples, recorded from task initiation to verbal completion confirmation, exhibited non-normal distributions (Shapiro-Wilk, $(p \leq 0.05)$). Puncture-related median durations were 13.82 seconds (V) and 17.57 seconds (MS). 
A Mann-Whitney U test showed no statistically significant difference between execution times with V modality and with MS modality ($U = 1316.0, p = 0.241$). 

\begin{figure}[!t]
    \centering
    \includegraphics[width=0.5\linewidth]{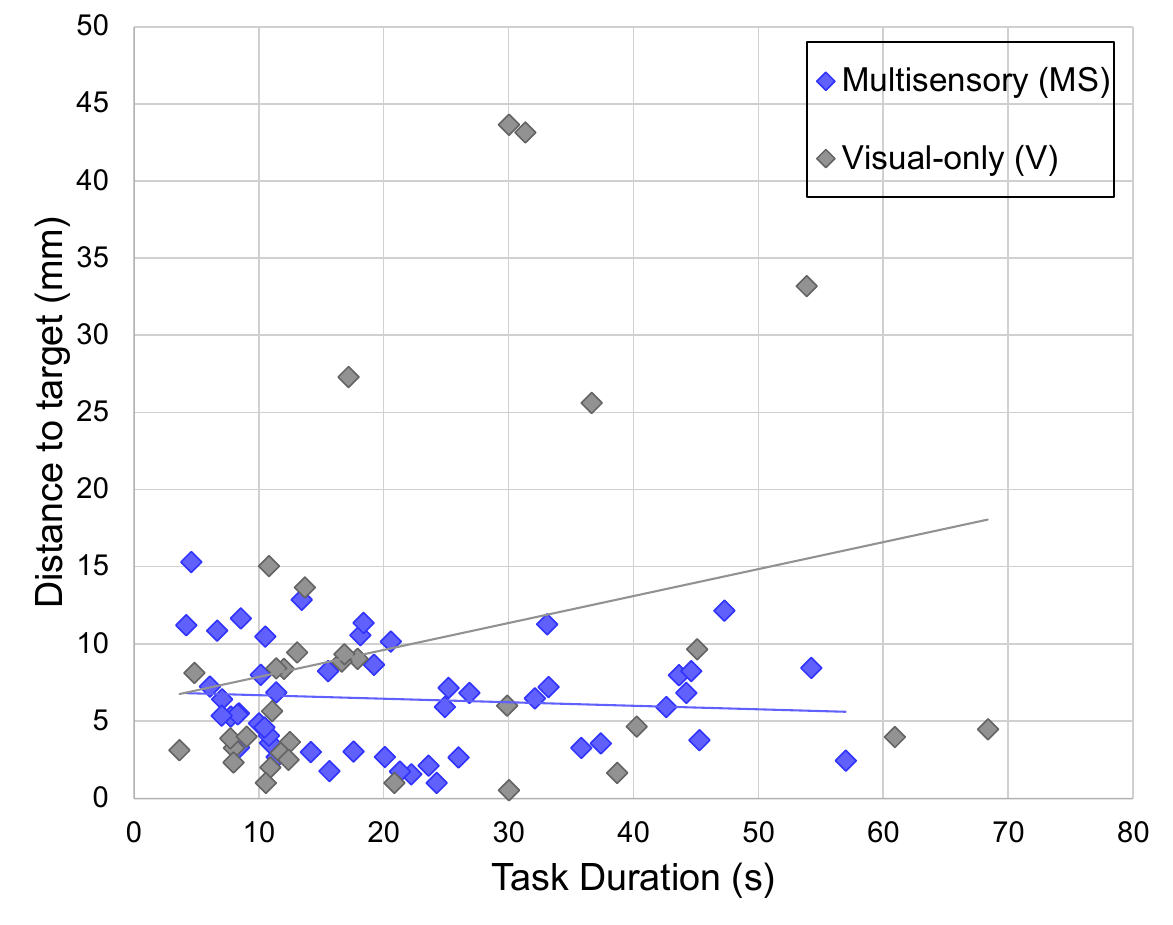}
    \caption{
        Modality-dependent time-accuracy relationships. Needle placement accuracy is measured by the distance of the needle tip position to the planned target point. The less distance represents better performance. Visual-only trials (gray) exhibit a positive correlation ($\rho = +0.31$), while multisensory trials (blue) show a negative trend ($\rho = -0.28$).
    }
    \label{fig:time-accuracy-correlation}
\end{figure}

The correlation between successful task duration and targeting accuracy (distance to target) is shown in Fig.~\ref{fig:time-accuracy-correlation}. 
A modality-dependent interaction was observed (Fisher’s $z = 3.12, p < 0.01$). Spearman’s rank correlations revealed opposite trends: 
\begin{itemize}
    \item \textbf{Visual-only}: Positive correlation ($\rho = +0.31, p < 0.05$), indicating that longer trials were weakly associated with larger errors (lower accuracy).  
    \item \textbf{Multisensory}: Negative correlation ($\rho = -0.28, p < 0.01$), suggesting that shorter trials tended to achieve higher accuracy.  
\end{itemize}

\subsection{Workload Assessment} 
A paired t-test revealed a statistically significant difference in perceived difficulty between conditions $(t = -4.02, p < 0.01)$, with lower difficulty ratings under MS compared to V. The analysis of the NASA-TLX questionnaire reported a relatively lower mental workload in the audiovisual condition (M = 10.00) compared to the visual-only condition (M = 12.67). While this difference was not statistically significant $(p = 0.059)$, the medium-to-large effect size $(\text{Cohen’s } d = 0.64)$ suggests a meaningful reduction. The 95\% confidence interval $[0.02, 5.32]$ further indicates that the audiovisual condition was generally perceived as less cognitively demanding.

\subsection{Experts' Subjective Feedback} 

Experts strongly favored this auditory feedback approach, noting that it enabled full reliance on sound during the navigation, particularly when visual cues were compromised by registration errors or OST-HMD instability. 

They highlighted that, unlike unimodal visual display, which often requires head and body movements (introducing a risk of misalignment), multisensory method offered millimeter-level precision and smooth control of the slender, highly sensitive needle.

Participants also emphasized its potential as a training tool for novices, who frequently have limited opportunities for supervised practice in PEA.

\section{Discussion}
\label{sect:discussion}

The proposed sonification approach goes far beyond simplistic, alarm-based auditory feedback, while it introduces a novel interactive design concept based on physical modeling synthesis, specifically using a multi-layered, dynamic 2D membrane that models the physical behavior of cardiac structures.
This modeling framework enables us to capture and sonify continuous spatial and temporal dynamics, such as the myocardium’s dynamic during diastole and systole, an aspect that to our knowledge has not been addressed in previous auditory feedback systems.
Unlike conventional binary or threshold-based sonification our system allows for expressive physically grounded audio rendering of the navigation process.
The parameters chosen (e.g., radius, tension, damping) reflect real physical characteristics and contribute to the perceptual intuitiveness of the sonification. 

Our results support the effectiveness of the proposed methodology across all evaluated metrics, with particularly improved task completion time and usability compared to previously reported needle guidance sonification approaches based on direct parameter mapping, which have been associated with increased cognitive demand and longer learning times~\cite{hansen2013auditory,matinfar2023sonification}.

\textbf{Navigation Safety:} Our findings demonstrate that multisensory feedback significantly outperformed visual-only modality for a safe navigation in PEA: MS modality significantly increased success rates and reduced both missed targets and critical failures. The stacked bar visualization (Fig.~\ref{fig:accuracy}) highlights the marked increase in success rate and the reduction in missed targets under the multisensory condition.
Notably, experts exhibited a 8\% \textit{Critical Failure} rate with both modalities, whereas novices had 8\% in V and 0\% with MS.
This can be explained that experts did not significantly rely on the navigational cues in both modalities due to their prior expertise. Instead novices, with no experience-related biases, completed the task without any critical failures with multisensory modality.
Even if the Chi-square test on the accuracy analysis conducted on expertise level subgroups doesn't indicate statical significance, the results may support the potential value of the multisensory guidance as training tool. Further studies with larger cohorts and repeated exposure to the system can substantiate these observations, and explore the interplay between expertise and multisensory integration.

\textbf{Needle Placement Accuracy:} The accuracy analysis of the final needle tip placement, critical for subsequent steps of the PEA procedure, yielded several insights based on the two accuracy measurements.
First, regarding the distance to the pericardial membrane, both modalities were comparably accurate, with no significant difference in median distance ($p=0.145$). However, the MS condition showed an 18\% reduction in variability (IQR), indicating more consistent and predictable outcomes. In this case, users were able to guide the needle into the pericardial sac while stopping in the proximity of the outer layer, suggesting that multisensory feedback enhanced perceptual awareness of the moving target.
Second, with respect to the distance from the predefined target points on the pericardial surface, MS guidance consistently reduced dispersion (lower MAD and IQR) with a tighter clustering effect around the target points, and decreased the number of outliers compared with V. Nevertheless, accuracy under this requirement remains uncertain for clinical translation, as targeting errors below 5 mm are recommended\cite{aryana2020percutaneous,koruth2011unusual}. It is also important to note that targeting error strongly depends on the initial needle-to-path alignment before navigation begins, since users had limited ability to adjust the trajectory once the needle was inserted into the foam chest patch. A deeper analysis that considers the entire task is therefore recommended.

\textbf{Execution Time:} The analysis, for successful placements only, revealed modality-specific differences in the speed–accuracy trade-off. Under V guidance, longer task durations were weakly associated with lower accuracy ($\rho=+0.31$), suggesting that additional time did not necessarily translate into more effective corrections and may instead reflect increased perceptual uncertainty. In contrast, the MS condition showed an opposite trend ($\rho=-0.28$), where longer trials tended to achieve higher accuracy. This finding supports the interpretation that multisensory feedback reduces perceptual confusion and facilitates more efficient and confident control, enabling users to succeed after a few corrective actions. The significance of the modality-dependent interaction ($p<0.01$) confirms that MS feedback fundamentally alters the speed–accuracy relationship compared to visual-only guidance, and indicates a potentially better learnability. Additional studies will be required to further investigate this relationship between execution time and accuracy, which is expected to be influenced by longer-term user adaptation and learning effects.

The analysis of the cognitive load using NASA-TLX and perceived difficulty complements the objective accuracy results, indicating that the multisensory condition not only improved performance but also reduced perceived cognitive effort.
Expert feedback further reinforces this interpretation.

These results should be interpreted in light of the primary goal of this study, which was to demonstrate the potential of a novel multisensory guidance paradigm. Importantly, the main outcome of the proposed approach lies in its ability to improve navigation safety, representing a significant step for this application domain. 

\textbf{Scalability into Clinical Settings:} While auditory feedback is the novel contribution of this work, the scalability of the proposed XR system into real clinical settings requires considerations. 
In particular, the dependence on accurate anatomical reconstruction, as well as on robust and efficient registration and tracking, may pose challenges to clinical adoption that requires to be investigated in future works.

A key strength of the framework is its modular architecture. Both the sonification and visualization components are driven by the same navigation parameters yet operate independently. This decoupling ensures that auditory feedback maintains high spatial accuracy regardless of the state of the visual channel, and the clinical scalability problem can be decomposed into two requirements of different criticality.

The primary requirement concerns the reliability and latency of the navigation parameters. These parameters describe the relative position of the needle tip with respect to the moving cardiac anatomy and are shared across both modules. However, since they are directly mapped to the sonification output, which is the component responsible for high-accuracy guidance, their accuracy and temporal consistency are of paramount importance.
While accurate and automatic myocardium reconstruction can be achieved from pre-operative 4DCTA data~\cite{bruns2022deep}, the robust reconstruction of thin structures such as the pericardium remains challenging, underscoring the need for targeted model development and validation.
4DCTA-based anatomical models are ECG-gated and therefore encode cardiac cycle information, which can be synchronized intraoperatively through ECG acquisition. 
Spatial registration requires support from an intraoperative imaging modality. A promising future direction is the use of intraoperative 3D ultrasound (US) as a bridge between the pre-operative and intraoperative domains: a 3D US volume is first registered to the 4DCTA using advanced deformable multimodal registration techniques~\cite{he2023robust, azampour2024anatomy}, establishing a common spatial reference. This reference then enables continuous intraoperative monitoring of cardiac motion through real-time registration of incoming 2D US frames~\cite{lei2024epicardium, wang2025eureg}, combined with myocardium segmentation~\cite{ying2025sam}. Together, these steps allow for ongoing refinement of the pre-operative anatomical models and dynamic correction of the navigation parameters throughout the procedure.
Regarding needle tracking, sensor-equipped catheters are already widely deployed in electrophysiology~\cite{hsu2022performance} and several platforms have received FDA clearance, making tracking accuracy and latency well-characterized for clinical translation.

The secondary requirement concerns the registration of the 4DCTA–derived anatomical models into the OST-HMD coordinate frame, ensuring that the visual overlay is correctly anchored to the patient. This challenge has been extensively explored in augmented reality surgical navigation research. Markerless registration based on anatomically meaningful landmarks currently represents the most suitable approach~\cite{kim2024automated}.

Regarding latency, no substantial increase is anticipated under clinical conditions compared to the phantom study. If intraoperative 2D US motion correction is enabled, the main additional contributor will be real-time image registration and segmentation, which must be carefully optimized to meet the temporal requirements of millimeter-scale navigation.

Finally, further studies are needed to quantify operating room sound masking effects. 

\section{Conclusion}
\label{sect:conclusion}

In this paper, we presented and evaluated a novel XR tool for intraprocedural guidance in PEA, where heart dynamics and need for high precision in needle positioning pose significant difficulties. The results highlight the potential of auditory feedback to support spatial awareness and decision-making during needle navigation, particularly in dynamic anatomical environments where visual cues alone may be insufficient. While the findings of the controlled experiment including expert users indicate potential benefits of multisensory guidance, they should be interpreted as preliminary. Further evaluation in more realistic and clinically relevant environments will be important to assess robustness, usability, and practical applicability. These results provide a foundation for future research on multisensory interaction and sonification in surgical guidance.

\section*{Acknowledgment}
This work was supported by MUSA – Multilayered Urban Sustainability Action – project, funded by the European Union – NextGenerationEU, under the National Recovery and Resilience Plan (NRRP) Mission 4 Component 2 Investment Line 1.5: Strenghtening of research structures and creation of R\&D “innovation ecosystems”, set up of “territorial leaders in R\&D”.

The authors would like to thank Leonardo Napoli and Mouzee S.r.l. (Torino, Italy) for kindly providing the physical phantom used in this study.

The authors declare that they have no conflicts of interest related to this work.

\section*{Supplementary Material}
\label{sec:suppl}
Supplementary Video S1 demonstrates the auditory mapping framework and the resulting sounds.\\
Supplementary Video S2 illustrates the user study conducted in the experimental set-up.

\bibliographystyle{unsrtnat}
\bibliography{references}  

\end{document}